\input{aipcheck}

\documentclass[
    ,final            
  ]
  {aipproc}

\usepackage{amsmath}
\usepackage{amssymb}

\layoutstyle{6x9}

\begin{document}

\title{The competition between the superconducting and the excitonic phases on doped Dirac electronic systems}

\classification{71.10.Fd,73.43.Nq,74.25.Dw,73.20.Mf}
\keywords{Dirac electrons,Superconductivity,Excitons,Quantum criticality}

\author{Lizardo H. C. M. Nunes}{
  address={Departamento de Ci\^encias Naturais, Universidade Federal de S\~ao Jo\~ao del Rei, 36301-000 S\~ao Jo\~ao del Rei, MG, Brazil}
}

\author{Ricardo L. S.  Farias}{
 address={Departamento de Ci\^encias Naturais, Universidade Federal de S\~ao Jo\~ao del Rei, 36301-000 S\~ao Jo\~ao del Rei, MG, Brazil}
}

\author{E. C. Marino}{
  address={Instituto de F\'{\i}sica, Universidade Federal do Rio de Janeiro, Caixa Postal 68528, Rio de Janeiro, RJ, 21941-972, Brazil}
}

\begin{abstract}
We investigate the competition between the superconducting and the excitonic phases on Dirac electrons on a bipartite planar lattice.
The conditions for the appearance of superconductivity or excitonic condensate are given by the minima of the free energy
and our results are explained.
\end{abstract}

\maketitle


\section{Introduction}
Several condensed matter systems
have been discovered in recent years whose active electrons have their kinematics
governed by the Dirac equation~~\cite{poly1,Neto_RMP09,cuprates,Direlpnic1}, 
 rather than by the Schr\"odinger equation. Their dispersion relation, accordingly, has the same form as that of a
relativistic particle. The reason for this unusual behavior of electrons whose speed is
at least two orders of magnitude less than the speed of light can be ascribed to a
particular influence of the lattice background on the electronic properties.

For many technological applications, it would be interesting if those systems were a semiconductor, instead of a semimetal.
Therefore, the possibility of an insulator state or the appearance of an excitonic gap due to short range electronic interactions
in a system of massless Dirac fermions 
has been the object of intense investigation~\cite{Araki2010,Weeks2010,Herbut2009}.
Nonetheless, early proposals that electron-electron interactions
could generate an electronic gap were investigated~\cite{Khveshchenko,Gorbar2002},
where the gap opening is an analogue of the ``chiral symmetry'' breaking process
that occurs in massless quantum electrodynamics (QED) in two dimensions~\cite{Appelquist1988}.
The excitonic interaction can be generated by different physical mechanisms. 
The Coulomb interaction, for instance, can generate it, precisely as it occurs in the case of the Hubbard model, 
where the on-site density-density repulsion term can be derived from the static Coulomb interaction.  
This is the type of excitonic instability that we consider in the present paper.

The fact that some of those condensed matter systems
become superconductors upon doping provides
a vast phenomenological motivation for the investigation
of the superconducting phase diagram of Dirac electrons systems.

Condensed matter systems with such competing interactions display phenomena that
are analogous to the onset of the chiral condensate and of color superconductivity in dense quark matter, which is an active subject of investigation~\cite{Alford2008}.
The phase transitions obtained in such systems can be compared with the ones occurring in analog condensed matter systems
with Dirac fermions.

\section{The effective potential}

Consider a system presenting a bipartite lattice formed of sublattices A and B, such as in graphene~\cite{Neto_RMP09}
or in the pnictides~\cite{Direlpnic1}.
Assuming that the interaction has two terms leading, respectively, to the superconducting and the excitonic instabilities,
our model Hamiltonian in the continuum limit reads~\cite{Nunes2012},
 \begin{equation}
H
=
\sum_{ \alpha }
\int \frac{ d^{ 2 } k }{ \left( 2 \pi \right)^{ 2 } }
\,
\Phi^{ \dagger }_{ \alpha } ( k )
\, \mathcal{A}_{ \alpha }  \,
\Phi_{ \alpha } ( k )
- \frac{ | \Delta |^2 }{ \lambda_{  { {sc}} } }
- \frac{ \sigma^2 }{ \lambda_{  { {exc}} } }
\, ,
\label{EqHCL}
\end{equation}
where  $ \alpha $ and $ v_{ \rm{ F } } $ denote the Dirac points and the Fermi velocity respectively,
$ t s_k = - v_{ \rm{ F } } \left(  k_{ x } - i k_{ y } \right) $,
$
\Phi^{ \dagger }_{ \alpha } ( k )
=
\left(
a^{\dagger}_{ \uparrow }( k )
\;
b^{\dagger}_{ \uparrow }( k )
\;
b_{ \downarrow }( -k )
\;
a_{ \downarrow }( -k )
\right)
$, 
which is written in terms of the electron creation operators of sublattices $ A $ and  $ B $,
and the matrix $ \mathcal{ A }_{ \bf k } $ is given by
\begin{equation}
\mathcal{A}_{ {\bf k }  }
=
\left(
 \begin{array}{cccc}
\mu + \sigma       &    - t s_{ k }             &        0                       &        \Delta              \\
 - t s^{ * }_{ k }     &    \mu  - \sigma      &         \Delta               &        0                     \\
0                          &    \Delta^{ * }          &          \sigma - \mu   &        t s^{ * }_{ k }    \\
\Delta^{ * }           &     0                        &          t s_{ k }           &       - ( \mu - \sigma )  \\
\end{array}
\right)
\label{EqMatrixA1}
\, ,
\end{equation}
with $ \sigma $ and $ \Delta $ denoting the excitonic and the superconducting order parameters.

The effective potential (free energy) per Dirac point reads
\begin{eqnarray}
V_{  { {eff}} }
& = &
\frac{ | \Delta |^2 }{ \lambda_{  { {sc}} } }
+
\frac{ \sigma^2 }{ \lambda_{  { {exc}} } }
\nonumber
\\
& &
-
T
\sum_{ n = -\infty }^{ \infty }
\int \frac{ d^2 k }{ (2 \pi )^2 }
\;
\log
\left\{
\frac
{
\prod_{ j = 1 }^{ 4 }
\left[
i \omega_n - E_j
\right]
}
{
\prod_{ j = 1 }^{ 4 }
\left[
i \omega_n - E_j( \sigma = 0, \Delta = 0 )
\right]
}
\right\}
\label{EqVeff}
,
\end{eqnarray}
    where $ \omega_n =  ( 2 n + 1 ) \pi T $ is the Matsubara frequency for fermions (with the Boltzmann constant
$k_B = 1 $) and the four $ E_ j $'s are
   \begin{equation}
E_j
=
\pm
\sqrt{
{
| \Delta |^2 +
\left[
  \sqrt{ \sigma^2 + ( v_{ \rm{ F} } k)^2 }
\pm \mu
\right]^2
}
}
\label{Ej}
\, .
\end{equation}
Notice that Eq.~(\ref{Ej}) is the typical dispersion relation observed in effective models for quantum chromodynamics,
where $ E^{ - }_ j  $ corresponds to the energy required to create a particle or a hole above the Fermi surface
and $ E^{ + }_ j  $ is the corresponding antiparticle term~\cite{Buballa2005}.
Moreover, the effective potential in Eq.~(\ref{EqVeff}) is the same obtained in~\cite{Ratti2004},
where the formation of chiral and diquark condensates involving two quark flavours in QCD was investigated.
In their case, the values for the quark-antiquark and quark-quark interactions strengths were set equal by construction,
via a Fierz transformation applied to the Nambu-Jona-Lasinio model,
which is not the case in the present paper,
since $  \lambda_{  { {sc}} } $ and $ \lambda_{  { {exc}} } $
are taken to be free parameters.

\section{Results and Conclusions}

The conditions for the appearance of superconductivity or excitonic condensate are given
by the nonzero solutions of $ \Delta $ or $\sigma $ which minimize the free energy,
those have been investigated already at zero temperature~\cite{Nunes2012}.
We have shown that, even if the excitonic interaction strength is larger than the superconducting interaction,
as the chemical potential increases, superconductivity eventually suppresses the excitonic order parameter,
which means that the system goes from an insulating state to a superconducting one
as charge carriers are added to the system.
Remarkably, similar results were obtained for a system for two-color and two-flavor QCD,
where chiral and diquark condensates were calculated~\cite{Fukushima2007}.

The model containing only the superconducting interaction at $ \mu = 0 $ has been previously investigated.
We have shown that there is a threshold for the superconducting interaction strength
which separates the normal and the superconducting states~\cite{Marino2006,Marino2007},
which is quantum critical point at $ \mu = 0 $, $ \lambda_{ c } $.
This critical interaction is expressed in terms of the ultraviolet cutoff $ \Lambda $,
namely, $ \lambda_{ c }  = 2 \pi v^{ 2 }_{ F } / \Lambda $~\cite{Marino2006},
where the Fermi velocity should be known for the actual physical system under investigation.
Therefore, the single free parameter of our model is $ \Lambda $
and the numerical results presented below are given
in terms of $ \Lambda $ and the dimensionless strength
$ \tilde{ \lambda }_{  { {sc}} } \equiv \lambda_{  { {sc}} } / \lambda_{ c } $.
Nonetheless, notice that the cutoff might be adjusted {\it a posteriori} through experimental data,
even if $ \Lambda $ is always provided by the lattice in condensed matter systems.

\begin{theacknowledgments}
This work has been supported in part by CNPq, FAPEMIG and FAPERJ. We would like to thank H. C. G. Caldas and A. L. Mota for
discussions on related matters.
\end{theacknowledgments}



\bibliographystyle{aipproc}   


\begin{thebibliography}{99}

\bibitem{poly1}
H. Takayama, Y. R. Lin-Liu, K. Maki,
Phys. Rev. B 21 (1980) 2388;

\bibitem{Neto_RMP09}
A. H. Castro Neto et al.,
Rev. Mod. Phys. 81 (2009) 109.

\bibitem{cuprates}
I. Affleck and J. B. Marston,
Phys. Rev. B 37 (1988) 3774;
Phys. Rev. B 39 (1989) 11538;
X-G. Wen and P. A. Lee,
Phys. Rev. Lett. 76 (1996) 503.

\bibitem{Direlpnic1}
C. M. S. da Concei\c c\~ao, M. B. Silva Neto and E. C. Marino,
Phys. Rev. Lett. 106 (2010) 117002.

\bibitem{Araki2010}
Y. Araki and T. Hatsuda
Phys. Rev. B 82 (2010) 121403(R).

\bibitem{Weeks2010}
C. Weeks and M. Franz,
Phys. Rev. B 81(2010) 085105.

\bibitem{Herbut2009}
V. Juri\~{c}i\'c, I. F. Herbut, and G. W. Semenoff,
Phys. Rev. B 80 (2009) 081405(R).

\bibitem{Khveshchenko}
D. V. Khveshchenko, Nucl. Phys. B 687 (2004) 323;
Phys. Rev. Lett. 87 (2001) 246802.

\bibitem{Gorbar2002}
E. V. Gorbar, V. P. Gusynin, V. A. Miransky, and I. A. Shovkovy,
Phys. Rev. B 66 (2002) 045108.

\bibitem{Appelquist1988}
T. Appelquist, D. Nash, and L. C. R. Wijewardhana,
Phys. Rev. Lett. 60 (1988) 2575.

\bibitem{Alford2008}
M. G. Alford et al.,
Rev. Mod. Phys. 80 (2008) 1455.

\bibitem{Nunes2012}
L. H.C. M. Nunes, R. L. S. Farias, and E. C. Marino
Phys. Lett. A 376 (2012) 779.

\bibitem{Buballa2005}
M. Buballa,
Phys. Rep. 407 (2005) 205.

\bibitem{Ratti2004}
C. Ratti and W. Weise,
Phys. Rev. D 70 (2004) 054013.

\bibitem{Fukushima2007}
K. Fukushima and K. Iida,
Phys. Rev. D 76 (2007) 054004.

\bibitem{Uchoa2005}
B. Uchoa, G. G. Cabrera and A. H. Castro Neto,
Phys. Rev. B {\bf 71} (2005) 184509.

\bibitem{Marino2006}
E. C. Marino and L. H. C. M. Nunes
Nucl. Phys. B {\bf 741} (2006) 404.

\bibitem{Marino2007}
E. C. Marino and L. H. C. M. Nunes,
Nucl. Phys. B {\bf 769} (2007) 275.

\end{thebibliography}

\end{document}